\documentclass[aps,prl,showkeys,showpacs,twocolumn,groupedaddress]{revtex4}

\usepackage{graphicx}

\begin{document}

\title{Are direct search experiments sensitive to all spin-independent WIMP
candidates?}

\author{F. Giuliani}
\email[]{franck@cii.fc.ul.pt}
\affiliation{Centro de F\'isica
Nuclear, Universidade de Lisboa, 1649-003 Lisboa, Portugal}

\date{April 19, 2005}

\begin{abstract}
The common analysis of direct searches for spin-independent Weakly
Interacting Massive Particles (WIMPs) assumes that a
spin-independent WIMP couples with the same strength with both
nucleons, \textit{i.e.} that the spin-independent interaction is
also fully isospin-independent. Though in a fully isospin-dependent
interaction scenario the spin-independent WIMP-nucleus cross section
is strongly quenched, the leading experiments are still sensitive
enough to set limits 1-2 orders of magnitude less stringent than
those traditionally presented. In the isospin-dependent scenario the
difference between the limits of CDMS-II and ZEPLIN-I is
significantly reduced. Here, a model-independent framework is
discussed and applied to obtain the current general
model-independent limits.
\end{abstract}


\maketitle

Direct searches for dark matter in the form of spin-independent
Weakly Interacting Massive Particles (WIMPs) are customarily
analyzed assuming that the WIMP couples with the same strength to
both protons and neutrons. This assumption, satisfied by various
theoretical candidates \cite{jung,Tovey,Savage}, corresponds to the
customary approximation Z $\approx$ N \cite{danai} (N is the neutron
number) which leads to the prejudice of inherent insensitivity of
the detectors to isospin-dependent spin-independent WIMPs. This
approximation, valid for light nuclei, breaks down for heavier ones,
like $^{127}$I ($\frac{N}{Z}=1.396$) and $^{131}$Xe
($\frac{N}{Z}=1.426$), which are used by various current and future
experiments. In this Letter, the effects of dropping this
approximation are discussed.

Direct searches are based on the detection of WIMP nonrelativistic
elastic scattering on nucleons, whose general effective
spin-independent lagrangian is \cite{Kurylov}

\begin{eqnarray}
{\cal L}=4\sqrt{2}G_{F}[\psi^{\dagger}\psi(g_{p}p^{\dagger} p +
g_{n} n^{\dagger} n)],
\label{lagrangian}
\end{eqnarray}

\noindent where $G_{F}$ is the Fermi constant, $\psi$, p and n are
the WIMP, proton and neutron two-component non-relativistic spinors,
and $g_{p,n}$ are the spin-independent WIMP-proton and WIMP-neutron
coupling strengths, respectively.

The general zero momentum transfer elastic scattering WIMP-nucleus
cross section $\sigma_{A}$ for a nucleus of mass number A resulting
from Eq. (\ref{lagrangian}) is:

\begin{equation}
\sigma_{A}=\frac{4}{\pi}G_{F}^{2}\mu_{A}^{2}(g_{p}Z+g_{n}N)^{2}
\label{basic}
\end{equation}

\noindent where $\mu_{A}$ is the WIMP-nucleus reduced mass. For
light nuclei $Z \approx N \approx \frac{A}{2}$, which substituted in
Eq. (\ref{basic}) yields

\begin{equation}
\sigma_{A}=\frac{4}{\pi}G_{F}^{2}\mu_{A}^{2}A^{2}\frac{(g_{p}+g_{n})^{2}}{4}
\label{sigmaASI},
\end{equation}

\noindent \textit{i.e.} light nuclei experiments are only sensitive
to $\frac{g_{p}+g_{n}}{2}$, which corresponds to the restriction of
isospin-independence: $g_{p}=g_{n}=\frac{(g_{p}+g_{n})}{2}$. Even
the experiments analyzing their data in a mixed-model framework,
\textit{e.g.} DAMA/NaI or HDMS \cite{danai,mixHdms}, retain this
restriction, which allows a four- instead of five-dimensional
parameter space.

\begin{figure*}
  \includegraphics[width=8 cm]{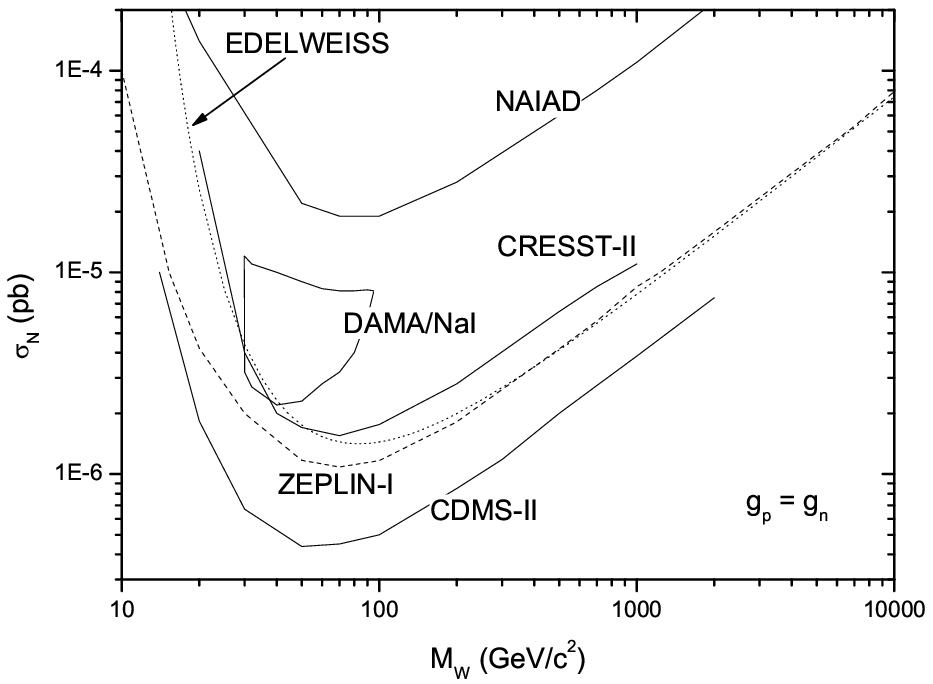}
  \includegraphics[width=8 cm]{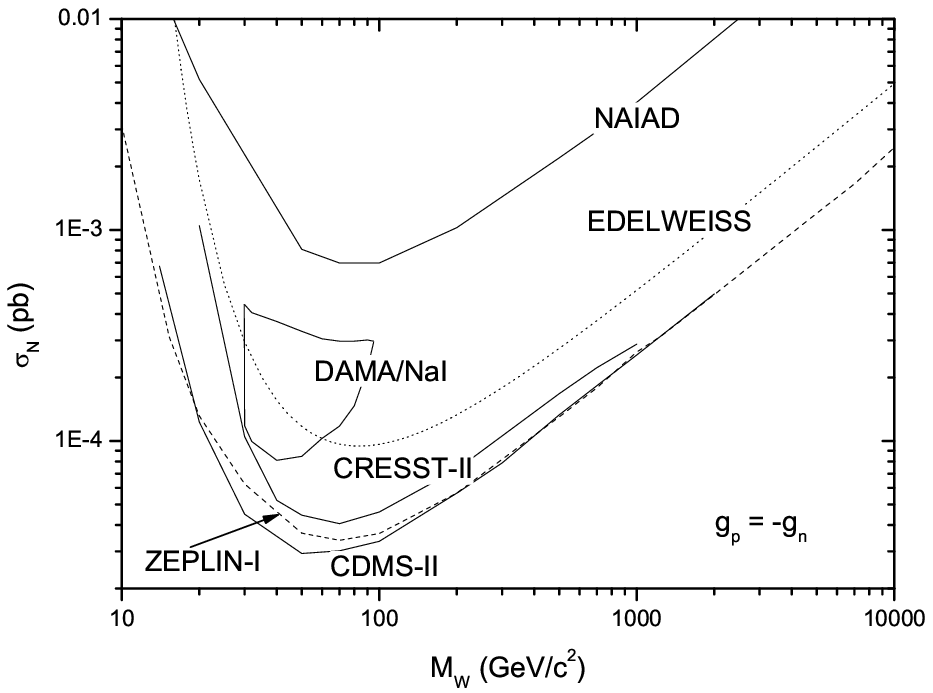}
  \caption{comparison between the exclusion limits normally presented by the experiments (left) and those for a fully isospin-dependent interaction (right).}\label{moddep}
\end{figure*}

If the interaction is fully isospin-dependent, \textit{i.e.}
$g_{p}=-g_{n}$, the two terms in Eq. (\ref{basic}) tend to suppress
the spin-independent $\sigma_{A}$, so that $^{127}$I would
effectively have only 21 nucleons and $^{131}$Xe 23, causing the
$^{127}I$ sensitivity to drop by a factor 36 and that of $^{131}Xe$
by a factor 32. The spin-independent sensitivity of light nuclei,
like F or Si would be negligible.

The impact of this model-dependence is illustrated in Fig.
\ref{moddep}, showing on the righthand side the case of a fully
isospin-dependent interaction. In this case the limits shift to
levels over an order of magnitude higher than the commonly reported
limits shown on the left side. Moreover, the relative positions of
the various exclusion plots change, as is understood in the light of
Table \ref{expos}, which reports the maximum value of N-Z for the
compositions of various leading experiments. Since Ge has N-Z
significantly lower than I or Xe, CDMS and EDELWEISS

\begin{table}[h]
\caption{\label{expos} maximum N-Z for the detector compositions of
the dark matter search experiments included in this Letter.}
\begin{ruledtabular}
\begin{tabular}{|c|c|c|c|}
\hline
material & experiments & max. N-Z & Refs. \\
\hline NaI & NAIAD, DAMA/NaI & 21 & \cite{naiad,danai} \\ \hline Xe
& ZEPLIN-I & 28 & \cite{ZEPLINI} \\ \hline
Ge & EDELWEISS, CDMS-II & 12 & \cite{edel,cdms} \\
\hline 
CaWO$_{4}$ & CRESST-II & 38 & \cite{papCRESSTII}\\ \hline
\end{tabular}
\end{ruledtabular}
\end{table}

\noindent are shifted more upwards than DAMA/NaI and ZEPLIN-I. The
fact that DAMA/NaI is shifted more upwards than ZEPLIN-I is a clear
consequence of Na's insensitivity to the (fully) isospin-dependent
interaction scenario, more than of the lower N-Z of I with respect
to Xe. The ability of CRESST-II to exclude DAMA/NaI improves, but in
spite of the high N-Z of W does not reach ZEPLIN-I nor CDMS.

The limits in the left-hand Fig. \ref{moddep} are obtained by
observing that for A=1 (proton or neutron) Eq. (\ref{sigmaASI})
becomes

\begin{equation}
\sigma_{N}=\frac{4}{\pi}G_{F}^{2}\mu_{p,n}^{2}\frac{(g_{p}+g_{n})^{2}}{4},
\label{asigmaSI}
\end{equation}

\noindent where $\sigma_{N}$ is the WIMP-nucleon cross section
\cite{Lewin}.

Substituting Eq. (\ref{sigmaASI}) in Eq. (\ref{asigmaSI}), yields
$\sigma_{N}=(\frac{\mu_{p,n}}{\mu_{A}})^{2}\frac{\sigma_{A}}{A^{2}}
\leq (\frac{\mu_{p,n}}{\mu_{A}})^{2}\frac{f_{A}
\sigma^{lim}_{A}}{A^{2}}$, f$_{A}$ being the fraction of isotope A
and $\sigma^{lim}_{A}$ the upper limit to $\sigma_{A}$ obtained by
attributing the entire rate to isotope A only. Introducing the new
symbol
$\sigma_{N}^{SI(A)}=(\frac{\mu_{p,n}}{\mu_{A}})^{2}\frac{\sigma^{lim}_{A}}{A^{2}}$
to indicate the overestimated limit obtained by attributing all the
observed rate to the isotope A, and recalling that
$\sum_{A}{f_{A}}=1$ results in \cite{Lewin}

\begin{equation}
\sigma_{N}\sum_{A}{[\frac{1}{\sigma_{N}^{SI(A)}}]}\leq1.
\label{mdsiexcl}
\end{equation}

In order to use the full Eq. (\ref{basic}), the following auxiliary
cross sections are introduced, in analogy with the spin-dependent
sector \cite{Tovey,FGprl}:

\begin{equation}
\left\{\begin{array}{l}
\sigma_{p}^{SI(A)}=(\frac{\mu_{p}}{\mu_{A}})^{2}\frac{\sigma_{A}^{lim}}{Z^{2}} \\
\sigma_{n}^{SI(A)}=(\frac{\mu_{n}}{\mu_{A}})^{2}\frac{\sigma_{A}^{lim}}{N^{2}}.
\end{array}\right .
\end{equation}

With these quantities, Eq. (\ref{basic}) becomes $\sigma_{A}=
\left(\sqrt{\frac{\sigma_{p}}{\sigma_{p}^{SI(A)}}} \pm
\sqrt{\frac{\sigma_{n}}{\sigma_{n}^{SI(A)}}}\right)^{2}\sigma_{A}^{lim}\leq
f_{A} \sigma_{A}^{lim}$, which implies:

\begin{equation}
\left\{\begin{array}{l}
\sum_{A}{\left(\sqrt{\frac{\sigma_{p}}{\sigma_{p}^{SI(A)}}} \pm
\sqrt{\frac{\sigma_{n}}{\sigma_{n}^{SI(A)}}}\right)^{2}}\leq1\\
\sum_{A}{\left(\frac{g_{p}}{\sqrt{\sigma_{p}^{SI(A)}}}+
\frac{g_{n}}{\sqrt{\sigma_{n}^{SI(A)}}}\right)^{2}}\leq
\frac{\pi}{4G_{F}^{2}\mu_{p}^{2}}
\end{array}\right .
\label{misiexcl}
\end{equation}

\begin{figure*}
\includegraphics[width=8 cm]{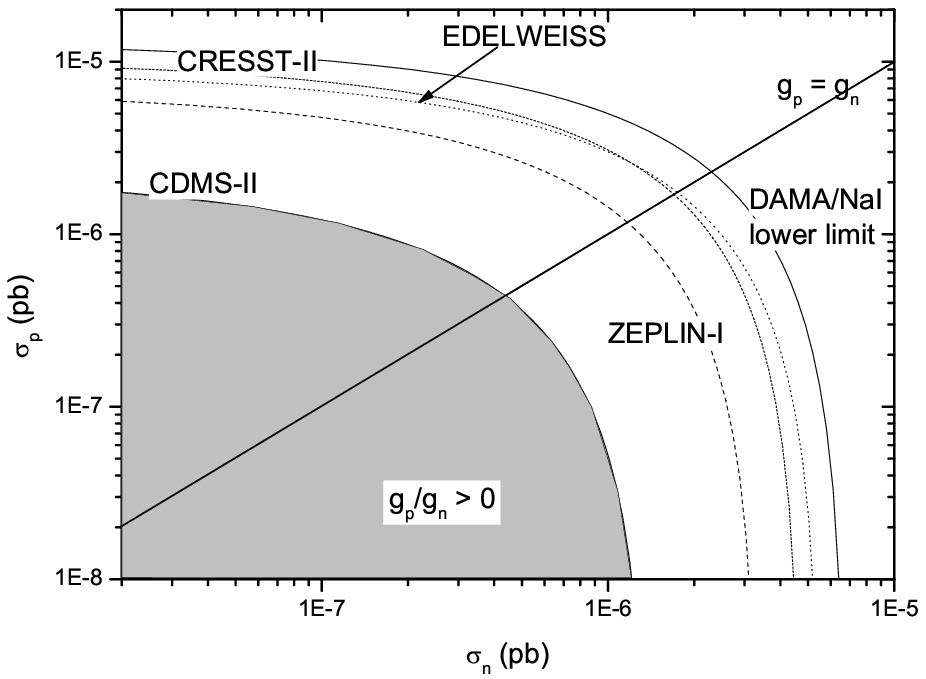}
\includegraphics[width=8 cm]{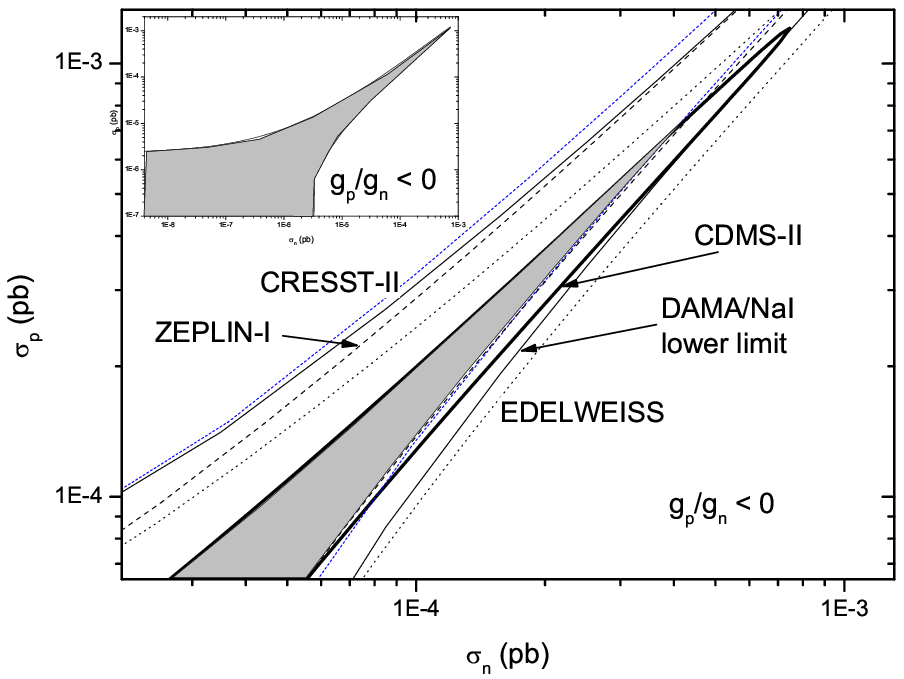}
\caption{cross section representation survey for
$\frac{g_{p}}{g_{n}}>0$ (left) and $\frac{g_{p}}{g_{n}}<0$ (right)
at M$_{W}=50$ GeV/c$^{2}$. The overall allowed regions are shaded.
The insert in the righthand figure shows the shape of the area
allowed by CDMS-II alone.}\label{sigsig}
\end{figure*}

\noindent where the sign of the sum inside the first parenthesis is
given by that of $\frac{g_{n}}{g_{p}}$, and the second is obtained
from the first through
$\sigma_{p,n}=\frac{4}{\pi}G_{F}^{2}\mu_{p,n}^{2}$ neglecting the
small difference between the proton and neutron mass. Extending the
terminology introduced in Ref. \cite{FGprd}, Eq. (\ref{misiexcl})
defines cross section and coupling strength representations for the
limits on spin-independent WIMPs. Since $\sigma_{p,n}^{SI(A)}$
depend upon M$_{W}$, the exclusion plots become three-dimensional,
which suggests, for illustrative purposes, to report bidimensional
cuts at constant M$_{W}$.

A cross section representation survey at M$_{W}=50$ GeV/c$^{2}$,
where CDMS-II reaches the most stringent limits, is shown in Fig.
\ref{sigsig}. In the left side $\frac{g_{p}}{g_{n}}>0$, and the
boundary of the region allowed by each experiment is a smoothly
decreasing convex curve. The region allowed by each experiment lies
within this curve, except for DAMA/NaI, whose allowed region lies
\textit{outside} its curve since only the lower limit curve is
shown. The straight line is the traditional condition of
isospin-independence, whose intersection point with the new
exclusion contour is the point of the traditional exclusion plot at
the chosen M$_{W}$. As evident, if $\frac{g_{p}}{g_{n}}>0$ the
intersection (shaded) of all allowed areas but DAMA/NaI coincides
with the CDMS-II area alone. For $\frac{g_{p}}{g_{n}}<0$ (righthand
Fig. \ref{sigsig}), instead, the region allowed by each experiment
(insert in the righthand Fig. \ref{sigsig}) has a generally finite
protuberance corresponding to $\frac{g_{p}}{g_{n}}\approx
-\frac{N}{Z}$, which is the least detectable interaction for its
composition: $g_{p}\approx-1.3g_{n}$ for Ge-based experiments, while
for I- and Xe-based is $g_{p}\approx-1.4g_{n}$ and for W
$g_{p}\approx-1.5g_{n}$. The protuberance of CDMS-II is cut by both
ZEPLIN-I and CRESST-II, being slightly tangent to the DAMA/NaI
contour, so that the couplings of the least constrained candidate
with M$_{W}=50$ GeV/c$^{2}$ are determined by the upper right point
of the shaded area in the righthand Fig. \ref{sigsig}. Again, the
region allowed by DAMA/NaI is the outside of its contour, which
entirely contains the shaded area allowed by all other experiments,
making the intersection of the DAMA/NaI region with the shaded area
empty. The emptyness of the intersection of the DAMA/NaI region with
that allowed by all other experiments for both
$\frac{g_{p}}{g_{n}}>0$ and $\frac{g_{p}}{g_{n}}<0$ translates the
incompatibility of the DAMA/NaI spin-independent candidate with the
observations of the remaining experiments.

\begin{figure}[h]
\includegraphics[width=9.25 cm]{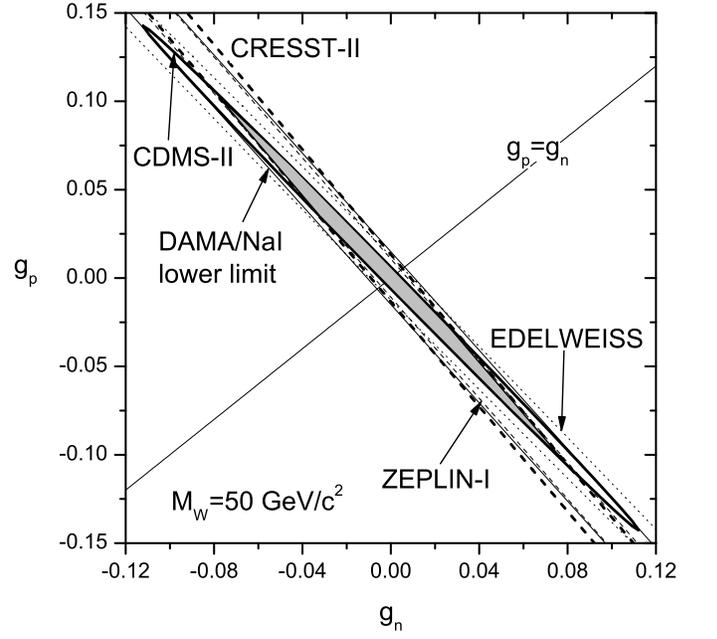}
\caption{spin-independent survey in the coupling strength
representation. 
The intersection of all experiments but DAMA/NaI is shaded. The
DAMA/NaI allowed region is the outside of the visible contour.
\label{ggsurv}}
\end{figure}

The coupling strength representation corresponding to Fig.
\ref{sigsig} is shown in Fig. \ref{ggsurv}. The parameters for Eq.
(\ref{misiexcl}) are obtained from the published traditional
spin-independent limits as follows: $\sigma_{N}$ is related to
$\sigma_{A}^{lim}$ by

\begin{eqnarray*}
\frac{1}{\sigma_{N}}=\sum_{A}{\frac{1}{\sigma_{N}^{SI(A)}}}=
\sum_{A}{\left(\frac{\mu_{A}}{\mu_{p,n}}\right)^{2}\frac{A^{2}}{\sigma^{lim}_{A}}}= \\
\sum_{A}{\left(\frac{\mu_{A}}{\mu_{p,n}}\right)^{2}f_{A}A^{2}}\frac{1}{\sigma^{lim}_{0}},
\end{eqnarray*}

\begin{figure*}
\includegraphics[width=7 cm]{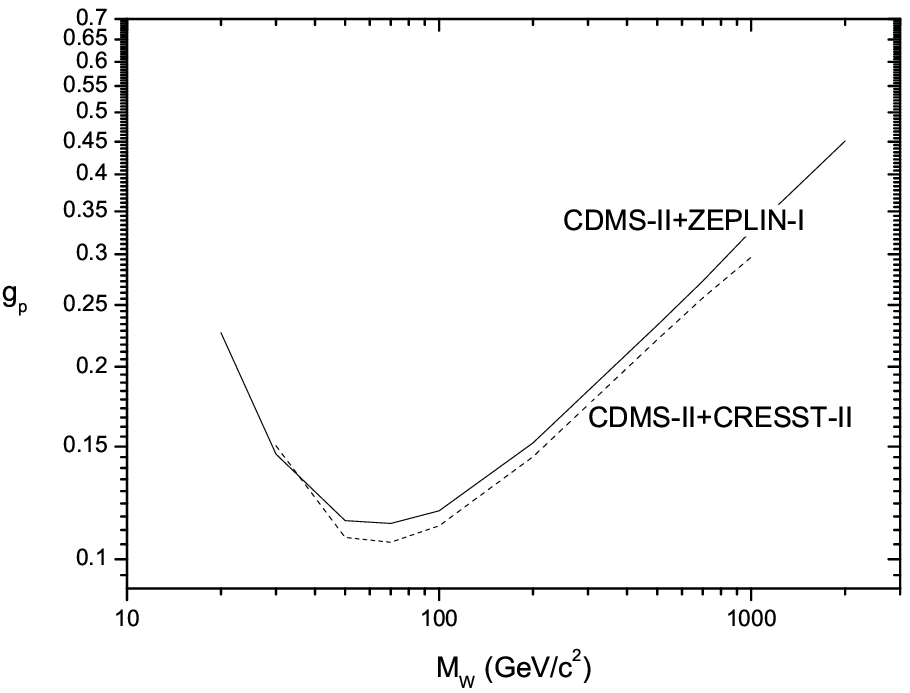}
\includegraphics[width=7 cm]{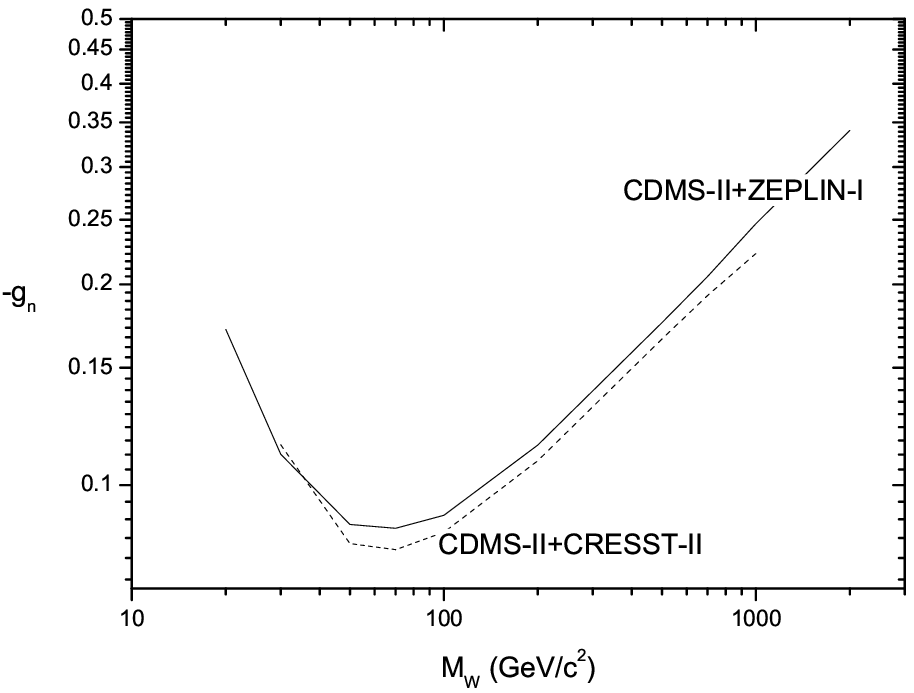}
\caption{overall spin-independent limits on $|g_{p}|$ (left) and
$|g_{n}|$ (right) from the intersection of CDMS-II and either
ZEPLIN-I or CRESST-II. In both cases the intersection points have
$\frac{g_{p}}{g_{n}}<0$. \label{inters}}
\end{figure*}

\noindent where $\sigma^{lim}_{0}=\frac{R}{J_{\psi}\rho}$ is the
"average" zero momentum transfer nuclear cross section, with R the
upper limit on WIMP-induced recoil rate, $J_{\psi}$ the incident
WIMP current and $\rho$ the total number density of the sensitive
material. Since
$\sum_{A}{(\frac{\mu_{A}}{\mu_{p,n}})^{2}f_{A}A^{2}}$ is easily
computable, $\sigma^{lim}_{0}$ can be calculated from the published
$\sigma_{N}$, and used as starting point for the model independent
re-analysis. Once $\sigma^{lim}_{0}$ is known, $\sigma^{lim}_{A}$ is
simply $\frac{\sigma^{lim}_{0}}{f_{A}}$, and Eq. (\ref{misiexcl})
can be applied. In the coupling strength representation, the
$g_{p}-g_{n}$ regions at constant M$_{W}$ allowed by each experiment
are the interior of ellipses, which degenerate to two parallel lines
for single nuclei experiments. The most striking difference with the
spin-dependent case is that, unlike the proton-to-neutron group spin
ratio, Z/N is always positive and less than 1, so that all
ellipses/bands have similar orientations. The overall
model-independent limits are still found by intersecting all
experimental conics, but the improvement obtained by intersection is
somewhat smaller. EDELWEISS contains entirely the CDMS ellipse,
because of the identical detector composition, and for this reason
is superseded by the latter. ZEPLIN-I and CRESST-II, instead, have a
slightly different orientation and cut part of the CDMS ellipse,
resulting in the shaded overall allowed region.

The overall combined limits from the intersection at M$_{W}=50$
GeV/c$^{2}$ are $|g_{p}|\leq 0.11$ and $|g_{n}|\leq 0.082$, or
$\sigma_{p}\leq 6.9 \times 10^{-4}$ pb and $\sigma_{n}\leq 4.0
\times 10^{-4}$ pb, about three orders of magnitude less stringent
than the usually assumed limits. The coupling strengths
corresponding to the traditional isospin-independent limits can be
found in Fig. \ref{ggsurv} by intersecting the $g_{p}=g_{n}$ line
with each experimental ellipse. The strong reduction in constraining
power is due to the large elongation of the ellipses, combined with
the small difference in their orientation. This translates
graphically the fact that each experiment has minimum sensitivity to
WIMPs for which $\frac{g_{p}}{g_{n}}\approx-\frac{N}{Z}$, while
$\frac{N}{Z}<2$ for practically all stable or long-lived isotopes.

Fig. \ref{inters} shows the locus of the least restrictive
intersection points of CDMS-II and either ZEPLIN-I or CRESST-II in
the coupling strength representation as a function of M$_{W}$. For
each WIMP mass, the intersection of two experiments is 2 pairs of
points symmetric with respect of the origin. Of these, the pair
farther from the origin has been selected, and within this pair the
point with $g_{p}>0$. These intersections have generally
$\frac{g_{p}}{g_{n}}<0$, because they correspond to small detector
sensitivities. As a consequence of the larger angle between the
CRESST-II and the CDMS ellipses, ZEPLIN-I, in spite of the lower
traditional limits, is generally superseded by CRESST-II when
determining the limits by intersection with CDMS.

Concluding, the leading experiments, being based on high mass number
isotopes, do not fulfill the traditional prejudice of the direct
spin-independent WIMP searches being only sensitive to
isospin-independent candidates. Though the experiments do not
constrain equally well all candidates with the same M$_{W}$, even
candidates whose coupling is primarily isospin-dependent
(\textit{i.e.} have $g_{p} \approx -g_{n}$) are constrained with 1-2
orders of magnitude weaker limits. The exclusion limits on the
candidate least constrained by the combination of the leading
experiments are 2-3 orders of magnitude less restrictive than
usually presented. The DAMA/NaI region obtained assuming the
standard halo model of Ref. \cite{Lewin} is excluded for all purely
spin-independent WIMP candidates satisfying Eq. (\ref{lagrangian}).

\begin{acknowledgments}
I wish to thank my colleagues of the SIMPLE collaboration for their
encouragement. I am supported by grant SFRH/BPD/13995/2003 of the
Portuguese Foundation for Science and Technology.
\end{acknowledgments}


\begin{thebibliography}{14}
\expandafter\ifx\csname
natexlab\endcsname\relax\def\natexlab#1{#1}\fi
\expandafter\ifx\csname bibnamefont\endcsname\relax
  \def\bibnamefont#1{#1}\fi
\expandafter\ifx\csname bibfnamefont\endcsname\relax
  \def\bibfnamefont#1{#1}\fi
\expandafter\ifx\csname citenamefont\endcsname\relax
  \def\citenamefont#1{#1}\fi
\expandafter\ifx\csname url\endcsname\relax
  \def\url#1{\texttt{#1}}\fi
\expandafter\ifx\csname urlprefix\endcsname\relax\def\urlprefix{URL
}\fi \providecommand{\bibinfo}[2]{#2}
\providecommand{\eprint}[2][]{\url{#2}}

\bibitem[{\citenamefont{Jungman et~al.}(1996)\citenamefont{Jungman,
  Kamionkowski, and Griest}}]{jung}
\bibinfo{author}{\bibfnamefont{G.}~\bibnamefont{Jungman}},
  \bibinfo{author}{\bibfnamefont{M.}~\bibnamefont{Kamionkowski}},
  \bibnamefont{and} \bibinfo{author}{\bibfnamefont{K.}~\bibnamefont{Griest}},
  \bibinfo{journal}{Phys. Rep.} \textbf{\bibinfo{volume}{267}},
  \bibinfo{pages}{195} (\bibinfo{year}{1996}).

\bibitem[{\citenamefont{Tovey et~al.}(2000)\citenamefont{Tovey, Gaitskell,
  Gondolo et~al.}}]{Tovey}
\bibinfo{author}{\bibfnamefont{D.~R.} \bibnamefont{Tovey}},
  \bibinfo{author}{\bibfnamefont{R.~J.} \bibnamefont{Gaitskell}},
  \bibinfo{author}{\bibfnamefont{P.}~\bibnamefont{Gondolo}},
  \bibnamefont{et~al.}, \bibinfo{journal}{Phys. Lett. B}
  \textbf{\bibinfo{volume}{488}}, \bibinfo{pages}{17} (\bibinfo{year}{2000}).

\bibitem[{\citenamefont{Savage et~al.}(2004)\citenamefont{Savage, Gondolo, and
  Freese}}]{Savage}
\bibinfo{author}{\bibfnamefont{C.}~\bibnamefont{Savage}},
  \bibinfo{author}{\bibfnamefont{P.}~\bibnamefont{Gondolo}}, \bibnamefont{and}
  \bibinfo{author}{\bibfnamefont{K.}~\bibnamefont{Freese}},
  \bibinfo{journal}{Phys. Rev. D} \textbf{\bibinfo{volume}{70}},
  \bibinfo{pages}{123513} (\bibinfo{year}{2004}).

\bibitem[{\citenamefont{Bernabei et~al.}(2001)\citenamefont{Bernabei, Amato,
  Belli et~al.}}]{danai}
\bibinfo{author}{\bibfnamefont{R.}~\bibnamefont{Bernabei}},
  \bibinfo{author}{\bibfnamefont{M.}~\bibnamefont{Amato}},
  \bibinfo{author}{\bibfnamefont{P.}~\bibnamefont{Belli}},
  \bibnamefont{et~al.}, \bibinfo{journal}{Phys. Lett. B}
  \textbf{\bibinfo{volume}{509}}, \bibinfo{pages}{197} (\bibinfo{year}{2001}).

\bibitem[{\citenamefont{Kurylov and Kamionkowski}(2004)}]{Kurylov}
\bibinfo{author}{\bibfnamefont{A.}~\bibnamefont{Kurylov}} \bibnamefont{and}
  \bibinfo{author}{\bibfnamefont{M.}~\bibnamefont{Kamionkowski}},
  \bibinfo{journal}{Phys. Rev. D} \textbf{\bibinfo{volume}{69}},
  \bibinfo{pages}{063503} (\bibinfo{year}{2004}).

\bibitem[{\citenamefont{Bednyakov and Klapdor-Kleingrothaus}()}]{mixHdms}
\bibinfo{author}{\bibfnamefont{V.~A.} \bibnamefont{Bednyakov}}
  \bibnamefont{and} \bibinfo{author}{\bibfnamefont{H.~V.}
  \bibnamefont{Klapdor-Kleingrothaus}}, \eprint{hep-ph/0504031}.

\bibitem[{\citenamefont{Ahmed et~al.}(2003)\citenamefont{Ahmed, Alner, Araujo
  et~al.}}]{naiad}
\bibinfo{author}{\bibfnamefont{B.}~\bibnamefont{Ahmed}},
  \bibinfo{author}{\bibfnamefont{G.~J.} \bibnamefont{Alner}},
  \bibinfo{author}{\bibfnamefont{H.}~\bibnamefont{Araujo}},
  \bibnamefont{et~al.}, \bibinfo{journal}{Astrop. Phys.}
  \textbf{\bibinfo{volume}{19}}, \bibinfo{pages}{691} (\bibinfo{year}{2003}).

\bibitem[{\citenamefont{Kudryavtsev and the Boulby Dark
  Matter~Collaboration}(2004)}]{ZEPLINI}
\bibinfo{author}{\bibfnamefont{V.~A.} \bibnamefont{Kudryavtsev}}
  \bibnamefont{and} \bibinfo{author}{\bibnamefont{the Boulby Dark
  Matter~Collaboration}}, in \emph{\bibinfo{booktitle}{Proc. of the 5th
  International Workshop on the Identification of Dark Matter}}
  (\bibinfo{address}{Edinburgh}, \bibinfo{year}{2004}).

\bibitem[{\citenamefont{Benoit et~al.}(2002)\citenamefont{Benoit, Bergé,
  Broniatowski et~al.}}]{edel}
\bibinfo{author}{\bibfnamefont{A.}~\bibnamefont{Benoit}},
  \bibinfo{author}{\bibfnamefont{L.}~\bibnamefont{Bergé}},
  \bibinfo{author}{\bibfnamefont{A.}~\bibnamefont{Broniatowski}},
  \bibnamefont{et~al.}, \bibinfo{journal}{Phys. Lett. B}
  \textbf{\bibinfo{volume}{545}}, \bibinfo{pages}{43} (\bibinfo{year}{2002}).

\bibitem[{\citenamefont{Akerib et~al.}(2004)\citenamefont{Akerib, Alvaro-Dean,
  Armel–Funkhouser et~al.}}]{cdms}
\bibinfo{author}{\bibfnamefont{D.~S.} \bibnamefont{Akerib}},
  \bibinfo{author}{\bibfnamefont{J.}~\bibnamefont{Alvaro-Dean}},
  \bibinfo{author}{\bibfnamefont{M.}~\bibnamefont{Armel–Funkhouser}},
  \bibnamefont{et~al.}, \bibinfo{journal}{Phys. Rev. Lett.}
  \textbf{\bibinfo{volume}{93}}, \bibinfo{pages}{211301}
  (\bibinfo{year}{2004}).

\bibitem[{\citenamefont{Angloher et~al.}(2005)\citenamefont{Angloher, Bucci,
  Christ et~al.}}]{papCRESSTII}
\bibinfo{author}{\bibfnamefont{G.}~\bibnamefont{Angloher}},
  \bibinfo{author}{\bibfnamefont{C.}~\bibnamefont{Bucci}},
  \bibinfo{author}{\bibfnamefont{P.}~\bibnamefont{Christ}},
  \bibnamefont{et~al.}, \bibinfo{journal}{Astrop. Phys.}
  \textbf{\bibinfo{volume}{23}}, \bibinfo{pages}{325} (\bibinfo{year}{2005}).

\bibitem[{\citenamefont{Lewin and Smith}(1996)}]{Lewin}
\bibinfo{author}{\bibfnamefont{J.~D.} \bibnamefont{Lewin}} \bibnamefont{and}
  \bibinfo{author}{\bibfnamefont{P.~F.} \bibnamefont{Smith}},
  \bibinfo{journal}{Astrop. Phys.} \textbf{\bibinfo{volume}{6}},
  \bibinfo{pages}{87} (\bibinfo{year}{1996}).

\bibitem[{\citenamefont{Giuliani}(2004)}]{FGprl}
\bibinfo{author}{\bibfnamefont{F.}~\bibnamefont{Giuliani}},
  \bibinfo{journal}{Phys. Rev. Lett.} \textbf{\bibinfo{volume}{93}},
  \bibinfo{pages}{161301} (\bibinfo{year}{2004}).

\bibitem[{\citenamefont{Giuliani and Girard}()}]{FGprd}
\bibinfo{author}{\bibfnamefont{F.}~\bibnamefont{Giuliani}} \bibnamefont{and}
  \bibinfo{author}{\bibfnamefont{T.}~\bibnamefont{Girard}},
  \eprint{hep-ph/0502232}.

\end{thebibliography}

\end{document}